# Hybrid Centrality Measures for Binary and Weighted Networks

**Alireza Abbasi, Liaquat Hossain**

Centre for Complex Systems Research, Faculty of Engineering and IT, University of Sydney, NSW 2006, Australia; alireza.abbasi@sydney.edu.au.

**Abstract** Existing centrality measures for social network analysis suggest the importance of an actor and give consideration to actor's given structural position in a network. These existing measures suggest specific attribute of an actor (i.e., popularity, accessibility, and brokerage behavior). In this study, we propose new hybrid centrality measures (i.e., *Degree-Degree, Degree-Closeness and Degree-Betweenness*), by combining existing measures (i.e., *degree, closeness and betweenness*) with a proposition to better understand the importance of actors in a given network. Generalized set of measures are also proposed for weighted networks. Our analysis of co-authorship networks dataset suggests significant correlation of our proposed new centrality measures (especially weighted networks) than traditional centrality measures with performance of the scholars. Thus, they are useful measures which can be used instead of traditional measures to show prominence of the actors in a network.

## 1 Introduction

Social network analysis (SNA) is the mapping and measuring of relationships and flows between nodes of the social network. SNA provides both a visual and a mathematical analysis of human-influenced relationships. The social environment can be expressed as patterns or regularities in relationships among interacting units [1]. Each social network can be represented as a graph made of nodes or actors (e.g. individuals, organizations, information) that are tied by one or more specific types of relations (e.g., financial exchange, trade, friends, and Web links). A link between any two nodes exists, if a relationship between those nodes exists. If the nodes represent people, a link means that those two people know each other in some way.

Measures of SNA, such as network centrality, have the potential to unfold existing informal network patterns and behavior that are not noticed before [2]. A method used to understand networks and their participants is to evaluate the location of actors within the network. These measures help determine the importance of a node in the network. Bavelas [3] was the pioneer who initially investigates formal properties of centrality as a relation between structural centrality and influence in group process. To quantify the importance of an actor in a social network, various centrality measures have been proposed over the years [4]. Freeman [5] defined centrality in terms of node degree centrality, betweenness centrality, and closeness, each having important implications on outcomes and processes.

While these defined measures are widely used to investigate the role and importance of networks but each one is useful based on especial cases, as discussed below:

(i) Degree centrality is simply the number of other nodes connected directly to a node. It is an indicator of an actor's communication activity and shows popularity of an actor;

(ii) Closeness centrality is the inverse of the sum of distances of a node to others ('farness'). A node in the nearest position to all others can most efficiently obtain information;

(iii) Betweenness centrality of a node is defined as the portion of the number of shortest paths that pass through the given node divided by the number of shortest path between any pair of nodes (regardless of passing through the given node) [6]. This indicates a node's potential control of communication within the network and highlights brokerage behavior of a node;

(iv) Eigenvector centrality is a measure of the importance of a node in a network. It assigns relative scores to all nodes in the network based on the principle that connections to high-scoring nodes contribute more to the score of the node in question than equal connections to low-scoring nodes. Bonacich [7] defines the centrality of a node as positive multiple of the sum of adjacent centralities.

For detail explanations and equations for the centrality measures please refer to [8].

In this study, we propose new centrality measures (i.e., *Degree-Degree, Degree-Closeness and Degree-Betweenness*), which combines existing measures (i.e., *degree, closeness and betweenness*) for improving our understanding of the importance of actors in a network. To show the significance of proposed new measure in evaluating actors' importance in the network, we first compare our proposed measures with a sample simple network and then we test it with a real co-authorship network having performance measure of nodes (scholars).

## 2 Hybrid Centrality Measures

To investigate the role and importance of nodes in a network, the traditional (popular) centrality measures could be applied in especial cases. By developing hybrid (combined) centrality measures, we are expecting to have a better understanding of importance of actors (nodes) in a network which can assist in exploring different characteristics and role of the actors in the network.

The proposed new measures work in combining (at least) two of the most popular and basic existing centrality measures of each actor. Thus, to achieve our goal, we propose three measures with an emphasis on degree, closeness and betweenness centralities of the direct neighbors of an actor. This will support in identifying the nodes which are central themselves and also connected to direct central nodes, which demonstrates strategic positions for controlling the network.

To define new hybrid centrality measures, we consider a network having centrality measures of each node as the attribute of the node. Then, we define hybrid centrality measures of a node as sum of centrality measure of all directly connected nodes. Thus, the *Degree-Degree (DD), Degree-Closeness (DC) and Degree-Betweenness (DB)* centralities of node *a* is given by:

$$DD(a) = \sum_{i=1}^{n} C_D(i) \quad , \quad DC(a) = \sum_{i=1}^{n} C_C(i) \quad , \quad DB(a) = \sum_{i=1}^{n} C_B(i)$$

Where *n* is the number of direct neighbors of node *a* (degree of node *a*) and $C_D(i)$ is the degree centrality measure, $C_C(i)$ is the closeness centrality measure and $C_B(i)$ is the betweenness centrality measure of node *i* (as a representation of direct neighbors of node *a*).

To have generalized measures, considering weighted networks which their links have different strengths, we can extend definitions by considering the weight of the links. Thus, the general hybrid centrality measures of node *a* are given by:

$$DD_w(a) = \sum_{i=1}^{n} [w(a,i) * C_D(i)] \quad , \quad DC_w(a) = \sum_{i=1}^{n} [w(a,i) * C_C(i)] \quad , \quad DB_w(a) = \sum_{i=1}^{n} [w(a,i) * C_B(i)]$$

Where *n* is the number of direct neighbors of node *a* and *w(a,i)* is the weight of the link between node *a* and its neighbors *i*.

Degree-Degree (DD) centrality indicates the actors who are connected better to more actors. It reflects the theory that connecting to more powerful actors will give you more power. So, it indicates the popularity of an actor based on popularity of its direct neighbors. Degree-Closeness (DC) centrality indicates not only an actors' power and influence on transmitting and controlling information but also efficiency for communication with others or efficiency in spreading information within the network. It indicates popularity and accessibility of an actor simultaneously. Also, Degree-Betweenness (DB) centrality indicates not only an actors' power and influence on transmitting and controlling information but also potential control of communication and information flow within the network. It shows popularity and brokerage attitude of an actor in the network simultaneously.

## 4 Applicability of new measures for analyzing nodes in networks

### 4.1 Simple examples

To compare our new proposed centrality measures and traditional centrality measures, we consider a simple network (Figure 1) and calculate nodes centrality measures (Table 1) and show the different ranks of the nodes based on each centrality measures in Table 2.

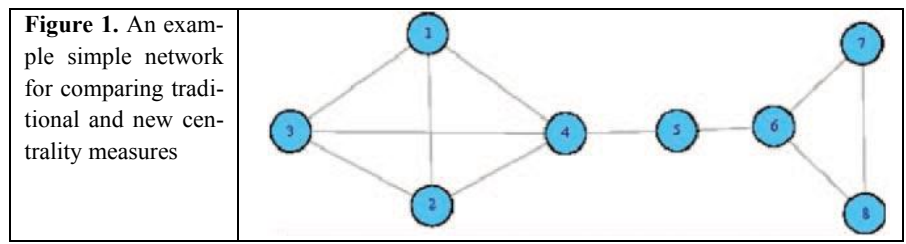

**Figure 1.** An example simple network for comparing traditional and new centrality measures

**Table 1.** Nodes' centrality measures for example network in Figure 1

| No | $C_D$ | $C_C$ | $C_B$ | $C_E$ | DD | DC | DB |
|---|---|---|---|---|---|---|---|
| 1 | .429 | .438 | 0 | .671 | 1.429 | 1.458 | 0.571 |
| 2 | .429 | .438 | 0 | .671 | 1.429 | 1.458 | 0.571 |
| 3 | .429 | .438 | 0 | .671 | 1.429 | 1.458 | 0.571 |
| 4 | .571 | .583 | .571 | .739 | 1.571 | 1.896 | 0.571 |
| 5 | .286 | .583 | .571 | .280 | 1.000 | 1.083 | 1.048 |
| 6 | .429 | .500 | .476 | .130 | 0.857 | 1.320 | 0.571 |
| 7 | .286 | .368 | 0 | .062 | 0.714 | 0.868 | 0.476 |
| 8 | .286 | .368 | 0 | .062 | 0.714 | 0.868 | 0.476 |

Table 2. Ranking nodes based on different centrality measures for network in Figure 1

| Rank | $C_D$ | $C_C$ | $C_B$ | $C_E$ | DD | DC | DB |
|---|---|---|---|---|---|---|---|
| 1 | 4 | 4,5 | 4,5 | 4 | 4 | 4 | 5 |
| 2 | 1,2,3,6 | 6 | 6 | 1,2,3 | 1,2,3 | 1,2,3 | 1,2,3,4,6 |
| 3 | 5,7,8 | 1,2,3 | 1,2,3,7,8 | 5 | 5 | 6 | 7,8 |
| 4 | | 7,8 | | 6 | 6 | 5 | |
| 5 | | | | 7,8 | 7,8 | 7,8 | |

As we expect the results and even ranks between traditional centrality measures are different except for eigenvector centrality ($C_E$, DD and almost DC). That is because the hybrid centralities can be considered as variants of eigenvector centrality.

## *4.2 A real co-authorship network*

Several studies have been shown the applicability of centrality measures for co-authorship networks for demonstrating how centrality measures are useful to reflect the performance of scholars (i.e., scholars' position within their co-authorship network) [8-10]. Here, also in another attempt, to assert the applicability of new hybrid centrality measures, we study a real co-authorship network having performance measure of actors (scholars) and their centrality measures, and test the correlation between centrality measures and performance measures.

**4.2.1 Data**

We analyzed the dataset which has been used in [8-9], publication list of five information schools: University of Pittsburgh, UC Berkeley, University of Maryland, University of Michigan, and Syracuse University. The data sources used are the school reports, which include the list of publications of researchers, DBLP, Google Scholar, and ACM portal. Citation data has been taken from Google Scholar and ACM Portal. Our data covered a period of five years (2001 to 2005), except for the University of Maryland iSchool, which had no data for the year 2002 in their report. We followed Google Scholars approach and did not differentiate between the different types of publications. After the cleansing of the publication data of the five iSchools, 2139 publications, 1806 authors, and 5310 co-authorships were finally available for our analysis.

### 4.2.2 Measuring Scholars' Performance

To assess the performance of scholars, many studies suggest quantifying scholars' publication activities (mainly citations count) as a good measure for the performance of scholars. Hirsch [11] introduced the h-index as a simple measure that combines in a simple way the quantity of publications and the quality of publications (i.e., number of citations). The h-index is defined as follows: "A scientist has an h-index of $h$, if $h$ of her $Np$ papers have at least $h$ citations each, and the other ($Np - h$) papers have at most $h$ citations each" [11]. In other words, a scholar with an index of $h$ has published $h$ papers, which have been cited at least $h$ times.

### 4.2.3 Results

The result of Spearman correlation rank test between centrality measures and scholars' performance (e.g., sum of citations and h-index) has been shown in Table 3. As it shows all traditional and new centrality measures are significantly correlated to performance measure except for eigenvector centrality and closeness which have weak or not significant correlations.

**Table 3.** Spearman correlation rank test between scholars' network centrality measures and their performance

| Centrality Measures (N=1806) | Scholars Performance | |
|---|---|---|
| | Sum_Cit. | h-index |
| $C_D$ | .332 ** | .311 ** |
| $C_C$ | - .012 | .052 * |
| $C_B$ | .388 ** | .501 ** |
| $C_E$ | .060 * | .041 |
| DD | .296 ** | .261 ** |
| DC | .303 ** | .295 ** |
| DB | .203 ** | .255 ** |
| $DD_W$ | .394 ** | .426 ** |
| $DC_W$ | .385 ** | .432 ** |
| $DB_W$ | .304 ** | .503 ** |

\*. Correlation is significant at the .05 level (2-tailed).
\*\*. Correlation is significant at the .01 level (2-tailed).

All new hybrid centrality measures of scholars have high positive significant association with their performance rather than traditional centrality measures. That is because the new measures combined two centrality measures' attributes and highlights the importance of the nodes in the network more than traditional ones.

The new centrality measures considering weighted links have higher correlation coefficients. This is due to taking into account scholar's repeated collaborations.

Another outcome of this result is that new centrality measure are different from eigenvector centrality and to support this we also applied non-parametric independent t-test (Mann-Whitney U test) to compare the distribution of eigenvector centrality measure between two groups (lower than mean of h-index and above mean) and it was not significant while the t-test was significant for new centrality measures. So, this also supports that new centrality measures are different from eigenvector centrality.

## 5 Conclusions

In this paper, we proposed a new class of hybrid centrality measures (i.e., DD, DC, DB). We illustrated similarities and dissimilarities with respect to the traditional (standard) measures considering a sample network and a real co-authorship network. Our analysis showed that they are good indicators of the importance of an actor in a social network by combing traditional centrality measures: degree of each node with degree, closeness and betweenness of its direct contacts for Degree-Degree, Degree-Closeness, Degree-Betweenness measures respectively. As each of them combines two different attributes (characteristics) of traditional measures, they could be a good extension of traditional centrality measures.

To demonstrate that the new measures are useful in practice to evaluate actors' importance in the network, we test it with having performance measures (e.g., sum of citations, h-index) of scholars. The results highlighted that Degree-Degree (DD), Degree-Closeness (DC) and Degree-Betweenness (DB) centralities have significant correlation with performance of the actors. Based on the results, we suggest that DD, DC and DB centralities of an actor are good measures to demonstrate the importance of an actor (e.g., performance, power, social influence) in a network.

It has been shown that in complex networks, Betweenness centrality of an existing node is a significantly better predictor of preferential attachment by new entrants than degree or closeness centrality [12]. We expect that the new proposed measure may be a better driver of attachment of new added nodes to the existing ones during the evolution of the network.

The computational complexity for calculating the proposed measure can be considered as one of the limitations of these new proposed measures which needs more research in future works. Also to generalize the applicability of the new hy-brid measures, it is needed to apply them in different (complex) networks in future works.

**Acknowledgments**   we appreciate Dr. Kenneth Chung's feedback on the earlier version of this work.

## References

1. Wasserman, S. and K. Faust, *Social network analysis: Methods and applications*. 1994: Cambridge Univ Press.
2. Brandes, U. and D. Fleischer. *Centrality measures based on current flow*. 2005: Springer.
3. Bavelas, A., *Communication patterns in task-oriented groups*. Journal of the Acoustical Society of America, 1950. **22**: p. 725-730.
4. Scott, J., *Social network analysis: a handbook*. 1991: Sage.
5. Freeman, L.C., *Centrality in social networks conceptual clarification*. Social Networks, 1979. **1**(3): p. 215-239.
6. Borgatti, S., *Centrality and AIDS*. Connections, 1995. **18**(1): p. 112-114.
7. Bonacich, P., *Factoring and weighting approaches to status scores and clique identification*. Journal of Mathematical Sociology, 1972. **2**(1): p. 113–120.
8. Abbasi, A., J. Altmann, and L. Hossain, *Identifying the Effects of Co-Authorship Networks on the Performance of Scholars: A Correlation and Regression Analysis of Performance Measures and Social Network Analysis Measures*. Journal of Informetrics, 2011. **5**(4): p. 594-607.
9. Abbasi, A. and J. Altmann. *On the Correlation between Research Performance and Social Network Analysis Measures Applied to Research Collaboration Networks*. in *Hawaii International Conference on System Sciences, Proceedings of the 44th Annual*. 2011. Waikoloa, HI: IEEE.
10. Yan, E. and Y. Ding, *Applying centrality measures to impact analysis: A coauthorship network analysis*. Journal of the American Society for Information Science and Technology, 2009. **60**(10): p. 2107-2118.
11. Hirsch, J., *An index to quantify an individual's scientific research output*. Proceedings of the National Academy of Sciences, 2005. **102**(46): p. 16569.
12. Abbasi, A., L. Hossain, and L. Leydesdorff, *Betweenness Centrality as a Driver of Preferential Attachment in the Evolution of Research Collaboration Networks*. Journal of Informetrics, 2012. **6**(1).